\documentclass{emulateapj}

\newcommand{\bi}{\begin{itemize}}
\newcommand{\ei}{\end{itemize}}
\newcommand{\nltt}{NLTT~11748}
\newcommand{\kms}{\ensuremath{{\rm km\,s}^{-1}}}
\newcommand{\dt}{\ensuremath{\Delta t_{\rm LT}}}

\shortauthors{Kaplan}
\shorttitle{Mass Constraints From Eclipse Timing in  Double WD Binaries}

\begin{document}
\title{Mass Constraints from Eclipse Timing in
  Double White Dwarf Binaries}

\author{David~L.~Kaplan\altaffilmark{1}}
\affil{KITP, Kohn Hall, University of
  California, Santa Barbara, CA 93106, USA; dkaplan@kitp.ucsb.edu}
\altaffiltext{1}{\textit{Hubble} Fellow}

\slugcomment{ApJ, in press}
\begin{abstract}
I demonstrate that an effect similar to the R\"omer delay, familiar
from timing radio pulsars, should be detectable in the first eclipsing double
white dwarf (WD) binary, \nltt.  By measuring the difference of the
time between the secondary and primary eclipses from one-half period (4.6\,s), one
can determine the physical size of the orbit and hence constrain the
masses of the individual WDs.  A measurement with uncertainty
$<0.1\,$s---possible with modern large telescopes---will
determine the individual masses to $\pm0.02M_\odot$ when combined with
good-quality ($<1\,\kms$) radial velocity data, although the
eccentricity must also be known to high accuracy ($\pm 10^{-3}$).
Mass constraints improve as $P^{-1/2}$ (where $P$ is the
orbital period), so this works best in wide binaries and should be
detectable even for non-degenerate stars, but such constraints require the mass ratio
to differ from one and undistorted orbits.
\end{abstract}

\keywords{binaries: eclipsing---stars: individual
  (NLTT~11748)---techniques: photometric---white dwarfs}

\section{Introduction}
Since the discovery of binary pulsars \citep{ht75}, precision timing
(typical uncertainties $<1\,\mu$s) has been used to derive a
variety of physical constraints \citep[see the discussion in][]{lk04}.
The arrival-time delay across the orbit (the R\"{o}mer
delay\footnote[2]{The R\"omer delay is named after O.~R\"omer, who used
  deviations from periodicity in the eclipses of Io by Jupiter to
  deduce a finite speed of light \citep{sterken05b}.}) immediately
gives the projected semimajor axis of the pulsar \citep{bt76}.  This,
especially when coupled with the relatively narrow mass distribution
of neutron stars \citep{tc99}, constrains the mass of the companion.

I contrast this with eclipse timing of planetary systems (typically
uncertainties are $\gtrsim$\,seconds).  Here, with
a mass ratio $\approx 10^{-3}$ the radial velocity curve gives a limit
on the mass of the companion planet.  With transiting systems, $\sin
i\approx 1$ and the mass of the planet is further 
constrained but not known uniquely \citep{cblm00}, although with knowledge
of the stellar parameters one can infer  the planetary mass and radius
\citep[e.g.,][]{bcg+01}.  If individual eclipses can be timed to high
precision (and here I mean both primary and secondary eclipses, i.e.,
transits and occultations), one can learn more about the system (e.g.,
\citealt{winn10}). Variations in the eclipse times can unveil the
presence of additional bodies in the system
\citep[e.g.,][]{assc05,hm05}, precession
\citep[e.g.,][]{me02}, and kinematics of the system
\citep{rafikov09}.  

With the recent discovery \citep{sks+10} of \object[NLTT
  11748]{\nltt}, an eclipsing double white dwarf (WD) binary with a
tight enough orbit that the binary will merge within a Hubble time, a whole
new series of questions may be asked.  The initial constraints are the
radial velocity amplitude of the lighter object (owing to the inverted
mass--radius relation of WDs, this object is the larger and
brighter member of the system) and the widths and depths of both
transit and occultation.  From this, assuming a cold C/O WD for the
heavier object, \citet{sks+10} were able to limit the masses and radii
of both objects, but could not determine unique constraints.
Measurement of spectral lines from the fainter object would determine
both masses uniquely, but this is challenging as the fainter object is
only $\approx 3.5$\% of the flux of the brighter.

A number of other close WD binaries have been discovered in the last
2 years (see Table~\ref{tab:wd} for those with undetermined
inclinations).  Most of them, like \nltt, appear to have a low-mass
($\lesssim 0.2\,M_{\odot}$) He WD in orbit with a more massive
(0.5--1.0\,$M_{\odot}$) C/O WD.  Such systems are of interest because
of their eventual evolution, with mass transfer brought on by
gravitational radiation \citep{nypzv01} and are presumed to be the
progenitors of highly variable objects: R~CrB stars, AM~CVn binaries,
and Type Ia supernovae \citep{it84,webbink84}.  Many of these binaries
are also of immediate interest as verification targets for the \textit{Laser Interferometer Space
  Antenna} (\textit{LISA}) mission \citep{nelemans09}.

\begin{deluxetable*}{l c c c c c c c l}
\tabletypesize{\footnotesize}
\tablewidth{0pt}
\tablecaption{Double WDs That Will Merge Within a Hubble
  Time And May Be Eclipsing\label{tab:wd}}
\tablehead{
\colhead{Object} & \colhead{$P_{\rm Orb}$} & \colhead{$K_2$} &
\colhead{$M_1$} & \colhead{$R_1$} & \colhead{$M_2$} &
\colhead{$R_2$}  & \colhead{\dt} & \colhead{Refs.}\\
 & \colhead{(hr)}& \colhead{(\kms)}& \colhead{($M_\odot$)} &
\colhead{($R_\odot$)} & \colhead{($M_\odot$)} & \colhead{($R_\odot$)}
& \colhead{(s)}
}
\startdata
SDSS~J1053+5200 & 1.02 & 265 & 0.26 & $>0.017$ & 0.20 & 0.04\phn & 0.2 & 1,2\\
SDSS~J1436+5010 & 1.10 & 347 & 0.45 & \phantom{$>$}0.014    & 0.22 & 0.04\phn & 0.7 &  1,2\\
SDSS~J0849+0445 & 1.89 & 367 & 0.65 & \phantom{$>$}0.012    & 0.17 & 0.05\phn &  2.0 &  1\\
WD~2331+290 & 4.08 & 156 & 0.39 & \phantom{$>$}0.015 & 0.32 & 0.016 &  0.5 & 3,4,5\\
SDSS~J1257+5428 & 4.55 & 323 & 0.92 & \phantom{$>$}0.009    & 0.15 & 0.04\phn &  4.7 &  6,7,8\\
\nltt      & 5.64 & 271 & 0.74 & \phantom{$>$}0.010    & 0.15 &
0.04\phn &  4.6 &  9,10\\
SDSS~J0822+2753 & 5.85 & 271 & 0.71 & \phantom{$>$}0.010    & 0.17 & 0.04\phn &  4.7 & 1\\
\enddata
\tablerefs{
1: \citet{kbapk09};
2: \citet{mbtl09};
3: \citet*{mdd95};
4: \citet{nnk+05};
5: \citet*{lbh05}
6: \citet{bmtl09};
7: \citet{kvk10};
8: \citet{mgs+10};
9: \citet{kv09};
10: \citet{sks+10}
}
\tablecomments{The values for $M_1$ assume an edge on orbit, i.e.,
  $\sin i=1$.  The values for $R_1$ were calculated from those values
  assuming a cold C/O WD \citep{ab98}.  Additional double-WD binaries
  exist, but they have inclinations known to exclude eclipses.}
\end{deluxetable*}

Of the 11 compact WD binaries known, only \nltt\ is known to be
eclipsing, but searches for the other sources are not uniformly
constraining and additional systems may yet be discovered.  The flux
ratios vary for the systems, and in some cases it may be easier to
directly measure the radial velocity curves for both members of the
binary.  Without two radial velocity curves, mass constraints are limited.  Such constraints
are invaluable in understanding the detailed formation histories and
expected evolution of these systems as well as in determining the
mass--radius relation from eclipse measurements.  Moreover, their use
as \textit{LISA} verification sources is improved by accurate
knowledge of the binary parameters.  In this {Letter}, I
discuss an effect that uses precision timing of the eclipses in such
double WD systems to help constrain the individual masses of the WDs.
This technique is known in other contexts, being common in radio
pulsar systems and planetary systems
\citep{kcn+07,hdd+10,ack+10}, although in the latter it is largely a nuisance
parameter and does not constrain the systems.  I discuss its
applicability to eclipsing double WD systems, the required
observational precision and the resulting accuracy.

\section{Light Travel Delay and Mass Constraints}
In a system with a circular orbit, one often speaks of the primary
and secondary eclipses as occurring exactly $1/2$ period
apart, but this is not the case. If the members of the binary are of
unequal mass the finite speed of light will cause an apparent shift in
the phase of the secondary eclipse from $P/2$, where $P$ is the period
of the binary \citep{loeb05,fabrycky10}.  This is similar to the
shifts in eclipse timing caused by a perturbing third body on a binary
system \citep{sd95,ddj+98,ddk+00,sterken05,lkk+09,qdl+09}, although here one
only requires two bodies and the frequency of the shift is known.

In the case of a planet with mass $m\ll M$ orbiting a star with mass $M$,
one has a primary eclipse when the planet is in front of the star.
The light is blocked at time $t=0$.  However, that light was emitted
earlier by the star, at time $t_1=0-a/c$, since it traveled a distance
$a$ (the semimajor axis).  For the secondary eclipse, the light is
emitted by the planet at time $t=P/2$ but is blocked $a/c$ later,
at $t_2=P/2+a/c$.  The difference of these times exceeds  $P/2$
by $\dt=t_2-t_1-P/2=2a/c$, the sought-after quantity.

For two finite masses, I consider two objects
orbiting their center of mass with period $P$, masses $M_1$ and $M_2$,
and semimajor axis $a$.  The total mass of the system is $M=M_1+M_2$,
and of course $4\pi^2a^3=P^2G M$; the first
object orbits at a radius $a_1=a(M_2/M)$ and the second object orbits
at a radius $a_2=a(M_1/M)$.

Near primary eclipse, the primary is at $[x,y]=[2\pi a_1 t/P, a_1]$
and the secondary is at $[-2\pi a_2 t/P,-a_2]$ at time $t$, with the
observer at $[0,-\infty]$.  I project the image of the two objects to
the barycenter at $y=0$.  This gives $x_{\rm B,1}=2\pi a_1
(t-a_1/c)/P$ and $x_{\rm B,2}=-2\pi a_2(t+a_2/c)/P$.  Eclipses occur
when these are equal, which has the solution $t_1=(a_1-a_2)/c$.  Near
secondary eclipse, the primary is at $[-2\pi a_1 (t-P/2)/P, a_1]$ and
the secondary is at $[2\pi a_2 (t-P/2)/P,-a_2]$.  Following the same
argument, eclipses occur when $t_2-P/2=(a_2-a_1)/c$.  So the eclipses
differ by $t_2-t_1=P/2+2(a_2-a_1)/c$.  The light-travel delay is again
$\dt=t_2-t_1-P/2$,
\begin{equation}
\dt=\left(\frac{2}{c}\right)\left(a_2-a_1\right)=\left(\frac{2 a}{c}\right)\left(
\frac{M_1-M_2}{M_1+M_2}\right),
\end{equation}
reaching  $2a/c$ when $M_2\ll M_1$, as expected.  From
Kepler's laws the mass function ${K_2^3 P}/2\pi G={M_1^3
  \sin^3i}M^{-2}$, where $K_2$ is the radial velocity amplitude of
object 2, and since it is a transiting system, $\sin i\approx 1$.
Substituting for $a$ and the masses, the time
delay in terms of observables and the mass ratio $q$ (where $q=M_2/M_1
\leq1$) is:
\begin{equation}
\dt=\frac{P K_2}{\pi c}(1-q).
\end{equation}

\subsection{Magnitude and Detectability}

The eclipse duration for a circular orbit is roughly $T\approx
2R_2P/(2\pi a)\approx 3\,$minutes \citep{winn10};  the duration of
ingress/egress $\tau$ is decreased by a factor of $R_1/2R_2\approx 8$,
$\tau\approx R_1P/(2\pi a)\approx 20\,$s (numerical results are
for \nltt); and ingress/egress are sharpest at inclinations of exactly
$90\degr$.  
The accuracy of the eclipse time determination largely depends  on 
the duration of the ingress/egress and the total number of photons
accumulated during ingress/egress, since the bottom of the eclipse is
only slightly curved (for the primary eclipse) if not flat (secondary
eclipse), and one can derive \citep[e.g.,][]{cye+08}
\begin{equation}
\sigma_{t_c}=\frac{\sigma}{\delta}\frac{\tau}{\sqrt{2 N_{\rm obs}}},
\end{equation}
where the eclipse has fractional depth $\delta$, each observation has
fractional uncertainty $\sigma$, and there are $N_{\rm obs}$
observations during $\tau$.  This holds in the limit that the noise is
uncorrelated \citep[cf.][]{cw09,skk10}, which should be true at the
level discussed here (photometric precision of $\gtrsim$\,mmag).  So,
$\sigma_{t_c}$ scales as the duration of ingress/egress divided by the
total signal-to-noise ratio accumulated during that portion of the
orbit.  Since $N_{\rm obs}\propto \tau$ for a constant observing
cadence, $\sigma_{t_c}\propto \sqrt{\tau}$.
A star with $V=16.5\,$mag like \nltt\ gives roughly $0.1\,{\rm
  photon\,s}^{-1}{\rm cm}^{-2}$ or $2\times 10^{5}$\,photons detected
during a 20 s ingress/egress with a 4 m telescope.  For an eclipse
depth of 5\% this means a precision on individual eclipse times of
$<1\,$s.

For  general binary systems, I can rewrite the expression
for \dt\ in terms of the primary mass $M_1$, $q$, and $P$ (eliminating
$K_2$):
\begin{equation}
\dt = \left(\frac{2 G M_1 P^2}{\pi^2 c^3}\right)^{1/3}
\frac{(1-q)}{(1+q)^{2/3}}\label{eqn:dt}
\end{equation}
For systems with primaries that are typical C/O WDs and with secondaries
that are He WDs, with $M_1=0.5-1 M_\odot$ and $q=1/6-1/2$, $\dt$ goes
from 0.5\,s to 7\,s for periods of 0.5--10\,hr (Table~\ref{tab:wd}).

Eclipse depths are functions of the radii and temperatures of the WDs,
as well as the bandpass, and are hard to predict with any generality.
The main factor that will change systematically for other double WD
systems is $\tau$ itself, which is $\propto R_1 P^{1/3}$.  This means the
signal-to-noise ratio (i.e., detectability) is
$\dt/\sigma_{t_c}\propto \sqrt{P/R_1}$, so the effect is easiest to see in
long-period binaries.  As the mass ratio approaches 1 the magnitude of
the delay decreases, limiting its utility, but in such systems it may
be easier to search for the second set of spectral lines (depending on
the temperatures of the objects).

This effect should also be present in partially degenerate (sdB+WD) or
non-degenerate binary systems.  The precision on the
eclipse times goes as $\sqrt{\tau}\propto \sqrt{R_{\rm small}}$.  If
the binary is wide enough that the larger star(s) are
undistorted by tides and hence the orbit remains strictly periodic, the
overall detectability $\dt/\sigma_{t_c}\propto \sqrt{P/R_{\rm small}}$
can actually increase over the double WD case I have been
considering.  Requiring $a\propto R_{\rm large}$ to minimize tidal
distortions, which scale as $(R_{\rm large}/a)^3$,  for a system
with a $0.2M_{\odot}$ M star one needs  periods of $\gtrsim 1\,$day to
have tidal effects that are as small as in the double WD systems.  For
the ingress/egress duration the radius of the smaller object $R_{\rm
  small}$ increases from $\sim 0.01R_{\odot}$ to $\sim 0.1R_{\odot}$
(for a $0.1M_\odot$ M star companion, for example), so if the period
increases by more than a factor of 10 then the wide system is more
easily detectable (the probability of eclipse does decrease as $R/a$,
though, and both primary and secondary eclipses must be seen).
However, the ephemeris must be known sufficiently well with tight
enough limits on (or measurements of) eccentricity so that the
light-travel delay is the only deviation from regularity (see below).

For \nltt, I recognize that $K_2$ is the radial velocity that was
measured since the heavier object is the fainter one.  So, $q\approx
0.15/0.71=0.21$, $K_2=271\,\kms$, and $P=5.64\,$hr, which give
$\dt=4.6\,$s.  \citet{sks+10} measured individual eclipse times to
$\sim 10$\,s, making it hard to detect an effect like this.  However,
this was using 45 s exposures on a 2 m telescope, while the
ingress/egress duration was only $\approx 20\,$s.  Increasing to 4\,m
or 8\,m will improve the S/N of individual exposures by a factor of
4--16, and using a cadence better matched to the orbit will help as
well, driving eclipse time uncertainties to $\lesssim 1\,$s (as
above).  This is sufficient to detect \dt; below I discuss how well
one can measure it and what constraints one can get from it.

\subsection{Comparison With Eccentricity}
The above discussion  considered circular orbits.  For eccentricity
$e>0$ the situation changes.  I note that the objects in
Table~\ref{tab:wd} have orbits that are consistent with circular orbits,
although quantitative limits for $e$ are not always given.  This
follows from their expected evolutionary histories,
where common-envelope evolution \citep{nvypz00} should have
circularized orbits.  Nonetheless, in case our understanding of
these systems is incorrect or some further evolution (such as
interaction with another body) may have caused non-zero eccentricity,
I consider the effect of a non-zero eccentricity on our detection of
\dt.

First, there
are changes to the expression for \dt\ \citep{fabrycky10}:
\begin{equation}
\dt=(\dt)_{e=0}\times\left( \frac{1-e^2}{1-e^2
  \sin^2 \omega}\right)\approx
(\dt)_{e=0}\times\left(1-e^2\cos^2\omega+{\cal O}(e^4)\right)
\end{equation}
where $\omega$ is the argument of pericenter.  As $(\dt)_{e=0}$ is
small to begin with, this is unlikely to be significant.  More
important, though, is that an additional term changes the relative
timing of the primary and secondary eclipses.  Following
\citet{winn10}:
\begin{equation}
\Delta t_{e}\approx \frac{P e}{\pi}\cos\omega
\end{equation}
and the primary eclipse also changes duration relative to the
secondary eclipse by the ratio $1+e\sin\omega$.  To compare $\dt$ and
$\Delta t_{e}$ means effectively comparing $K_2(1-q)/c$ and
$e\cos\omega$.  For $K_2\sim 300\,\kms$, this means that one is
sensitive to $e\sim 10^{-3}$ (although $\omega$ is poorly determined
for low $e$).  If it can be asserted for some independent reason
(i.e., evolutionary assumptions) that $e\ll 10^{-3}$ then one can
treat any measured $\Delta t$ as coming from light-travel delay.  But
if not one must be more careful.

Fortunately, eccentricity can be constrained from the radial
velocities.  Adopting the small-$e$ limit as in \citet{lcw+01},
$N_{\rm RV}$ spectra can limit the eccentricity to $\sigma_e\approx
2\sigma_v/K_2\sqrt{N_{\rm RV}}$, where $\sigma_v$ is the precision of
the individual velocity measurements (see also \citealt{gw07}).  With
$>100$ observations with $<1\,\kms$ precision the eccentricity can be
limited (independent of $\omega$) to $\ll 10^{-3}$, and hence can identify
whether any measured time delay has a contribution from an eccentric
orbit.  This requires dedicated radial velocity measurements over one
or more full orbits, but is achievable with current instrumentation.
At this level one must also account for additional effects such as
light-travel delay in the {spectroscopic} analysis
\citep{za07}.  Tidal distortions can also mimic eccentricity in radial
velocity fits \citep{eaton08}, but these can be identified
photometrically and are expected to be quite small, $\sim 10^{-4}$,
except in the most compact systems.

\subsection{Mass Constraints}
Equation~(\ref{eqn:dt}) gives  an independent constraint on the
mass ratio $q$, which helps  break the degeneracy in the mass
function to measure the masses of the stars individually.  
For the individual masses
\begin{eqnarray}
M_1 & = & \frac{K_2}{2\pi G P}\left(2 P K_2-\dt \pi
c\right)^2 \nonumber \\
M_2 & = & \left(2 P K_2-\dt \pi
c\right)^2\left(\frac{K_2}{2\pi G P}-\frac{\dt  c}{2 G P^2}\right).
\end{eqnarray}

\begin{figure}
\plotone{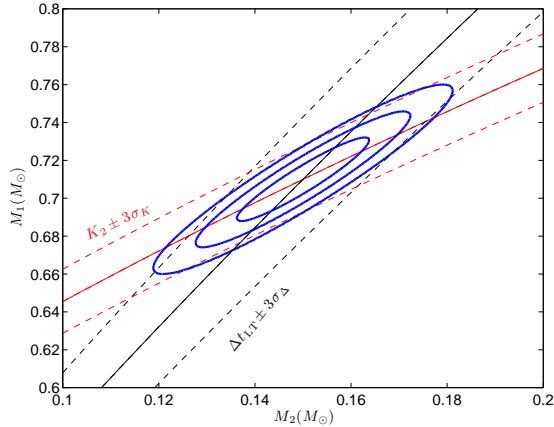}
\caption{Constraints on individual masses from a single radial velocity
  constraint $K_2$ and a light-travel delay \dt.  I show the
  constraints from each measurement individually (along with
  3$\sigma$ ranges) as the diagonal lines.  The contours show 1$\sigma$, 2$\sigma$,
  and 3$\sigma$ joint confidence contours on $M_1$ and $M_2$; their
  covariance is apparent.  This system has parameters similar to those
  of \nltt, and I assumed $\sigma_K=1\,\kms$ and
  $\sigma_\Delta=50\,$ms, which is rather optimistic.
\label{fig:mass}
}
\end{figure}

Assume that I measure $K_2 \pm \sigma_K$ and $\dt \pm \sigma_\Delta$ ($P$
is typically known to much higher precision); I also assumed $e=0$.
How well can I determine the individual masses?  I know $q$ to:
\begin{equation}
\sigma_q^2=\frac{\pi^2 c^2}{P^2 K_2^4}\left(K_2^2 \sigma_\Delta^2 + \dt^2 \sigma_K^2\right).
\end{equation}
I now wish to see with what precision I can estimate the masses from
the observations.  Doing standard error propagation, 
\begin{eqnarray}
\frac{\partial M_1}{\partial K_2} & = & \left(\frac{P M^2}{2\pi G}\right)^{1/3}\frac{5+q}{(1+q)}
\nonumber \\
\left|\frac{\partial M_1}{\partial \dt}\right| & = & \left(\frac{4 \pi^2 M^2c^3}{P^2G}\right)^{1/3}\frac{1}{(1+q)}.
\end{eqnarray}
These are the contributions of the $\sigma_K$ and $\sigma_\Delta$ to
the uncertainty on the mass, i.e.,
$\sigma_M^2=\sigma_K^2\left|\partial M/\partial
K\right|^2+\sigma_\Delta^2\left|\partial M/\partial \dt\right|^2$.
For the two terms to be comparable requires
$\sigma_\Delta=((q+5)P/2\pi c)\sigma_K\approx 10P_{\rm
  hr}\sigma_{K,{\rm kms}}\,{\rm ms}$ (where $P_{\rm hr}$ is the period in
hr and $\sigma_{K,\rm kms}$ is the uncertainty on $K_2$ in \kms).
With those, I would have $\sigma_{M_1}\sim 0.01 M_\odot$ for periods
$P\gtrsim 1\,$hr and mass ratios $q\sim 0.25$. The constraint on $M_2$
is similar.  I illustrate this in Figure~\ref{fig:mass}, where I
show mass constraints on \nltt\ from hypothetical time-delay
measurements.

However, it is likely that the uncertainty from the radial velocity
amplitude will be considerably less than that from \dt: individual
velocity measurements can easily have uncertainties of a few
\kms\ with a large telescope, and combining enough of them to give a
meaningful constraint on the eccentricity will likewise end up with
$\sigma_K=\sigma_v/\sqrt{N_{\rm RV}}< 1\,\kms$.  Getting a comparable
constraint on the time delay seems implausible: with individual times
measured to $\sim 1\,$s, $>10^4$ eclipses are needed to get
$\sigma_\Delta<10\,$ms, and only one time delay is measurable per
orbit.  This means one needs significantly higher signal-to-noise
per observation than the several hundred I have been assuming here or
that I will be limited by $\sigma_\Delta$.  In this case, the precision
on $M_1$ improves as $\sigma_{M_1}\sim P^{-2/3}\sigma_\Delta$;
including the difficulty in detecting the delay $\sigma_\Delta\sim
P^{1/6}$,  the determination of $M_1$ improves as
$P^{-1/2}$.  The only ways that long periods are penalized are in
terms of observing strategy, as for long periods the time to get
enough eclipses measured will grow long as well, and for the
probability of detecting an eclipse in the first place since that
decreases as $1/a$.

As shown in Figure~\ref{fig:mass}, the joint probability distribution
for $(M_1,M_2)$ is strongly correlated between $M_1$ and $M_2$, with a
much stronger constraint on $M_2-M_1$ than on $M_2+M_1$.  However,
this would even be true---although to a lesser degree---if
the other radial velocity amplitude $K_1$ were measured, especially if it
were at lower precision because it is much fainter.  It is
straightforward although tedious to compute the linear combinations of
the masses that minimize/maximize the variance which would be preferred
for fitting.  While not perfect, with this constraint one would have a
much  better picture of the system.

\section{Conclusions}
Motivated by the recent discovery of \nltt, the first eclipsing double
WD binary, I have examined a phenomenon that affects precision eclipse
timing of such a system.  With knowledge of the individual masses, one
can put much stronger constraints on the radii of the two WDs, the evolutionary history of the system, and its expected
outcome, not to mention WD atmosphere models and models for
the interiors of He WDs \citep[e.g.,][]{pach07,sba10}.  This effect, a
delay between perfect phasing of primary and secondary eclipses, can be
used to constrain the individual masses of the binary, something
difficult to do otherwise.

I find that the light-travel delay should be detectable in the case of
\nltt, and possibly in some other similar binary systems should they
prove to have both primary and secondary eclipses: long periods are
favored both for detecting \dt\ and for using it to constrain the
masses, although long periods do not favor detecting eclipses to begin
with.  The constraints on the individual masses can approach
$\pm0.01M_{\odot}$ for plausible data-sets, and will likely be limited
by the precision of the eclipse timing, suggesting that an intensive
timing effort on large telescopes is worthwhile.  Detection of a
second radial velocity amplitude would over constrain the system,
leading to even tighter determinations of the masses.  In
non-degenerate eclipsing binaries, such as those that \textit{Kepler}
may discover, the delay should also be detectable for
systems with orbital periods of greater than a few days, although it
requires that the mass ratio differs from one and that no other unmodeled
orbit variations be present to a high degree of confidence.

The orbits of the binary members can also be perturbed by other bodies
in the systems, either on shorter (planets or other small bodies in
close orbits) or longer timescales (a distant body).  In both cases,
perturbations in transit timing may be visible (see \citealt{assc05}
for a detailed discussion).  Given the wide variety in possible
situations it is out of the scope of this Letter to consider, but any
perceived variation in transit timing must be compared against the
possible presence of additional bodies.  There could also be effects
that alter the perceived primary versus secondary eclipse times
without altering the orbit, such as hot spots due to irradiation
\citep{kcn+07,ack+10} or accretion.  For the former, I note that the
incoming radiation in the double WD systems is typically very small,
$\sim 10^{-3}$ of the outgoing radiation.  As for accretion, both WDs are
well inside their Roche lobes and so none is expected.

I find the fortunate coincidence that \nltt, the one object known to
be eclipsing, also has the binary parameters that lead to the highest
value of \dt\ among similar double WD binaries.  Hopefully, with
dedicated observing  this effect will be detected
and can constrain the \nltt\ system even more than is possible today.

\acknowledgements I thank the anonymous referee, as well as L.~Bildsten,
T.~Marsh, J.~Winn, E.~Agol, D.~Fabrycky, S.~Gaudi, M.~van~Adelsberg,
and R.~Cooper for helpful discussions.  DLK was supported by NASA
through Hubble Fellowship Grant \#01207.01-A awarded by the STScI
which is operated by AURA, Inc., for NASA, under contract NAS 5-26555.
This work was supported by the NSF under grants PHY 05-51164 and AST
07-07633.



\end{document}